\documentstyle[11pt,paspconf]{article}
\def\etal{{\it et~al.\ }}

\def\ie{{\it i.e.,~}}
\def\de{\partial}
\def\ltsima{$\; \buildrel < \over \sim \;$}
\def\simlt{\lower.5ex\hbox{\ltsima}}
\def\gtsima{$\; \buildrel > \over \sim \;$}
\def\simgt{\lower.5ex\hbox{\gtsima}}

\begin{document}

\title{Acoustic and Thermoreactive Instabilities in a Photoionized Multiphase
Medium}
\author{A. Ferrara and E. Corbelli}
\affil{Osservatorio Astrofisico di Arcetri, Firenze, Italy}

\begin{abstract}
We study thermoreactive and acoustic
instabilities in a diffuse gas, photoionized and heated by a radiation
field.
The analysis of the thermal instability by Field (1965) is
extended to include the effects of the hydrogen recombination reaction,
which, in general, is found to be a stabilizing agent for the condensation
mode. This effect is stronger when the mean
photon energy is not much larger than the hydrogen ionization energy.
In addition, there are thermoreactive unstable equilibria
for which an isobaric transition to a stable phase is not possible
and the system evolves toward  a dynamic state characterized
by large amplitude, nonlinear periodic oscillations of temperature,
density and hydrogen ionization fraction.

Acoustic waves are found to be unstable for some temperatures of
both the cold and the warm phase
of the ISM, also in regions where the gas is  {\it thermally stable},
independently of the mean photon energy.
\end{abstract}


\section{Introduction}
Photoionization and radiative energy input processes are important in many
different interstellar and intergalactic environments. Some examples are
provided by Ly$\alpha$ clouds (Ikeuchi \& Turner 1991, Charlton \etal 1993,
Kulkarni \& Fall 1993), outer disks of spiral galaxies (Corbelli \& Salpeter
1993, Maloney 1993, Dove \& Shull 1993), broad line emitting regions
of AGNs (Kwan \& Krolik 1981, Collin-Souffrin 1990), High Velocity Clouds
(HVCs) in the Galactic halo (Songaila etal 1989, Ferrara \& Field 1994,
McKee 1994) and the extended Galactic free electron layer (Reynolds 1993).

One of the interesting characteristics of  photoionized objects is that, due
to ongoing thermal instabilities, they often develop a multiphase structure.
An example of a two-phase medium in which  cold neutral medium
coexists in pressure equilibrium with warm ionized medium is shown in
Fig. 1, taken from Ferrara \& Field (1994), which describes the ionization
state of HVCs. A HVC in the Galactic halo is immersed in the extragalactic
background radiation and the external layers of the cloud become
ionized. For low values of the pressure of the external confining medium
(for example, hot gas from a fountain) the cloud is optically thin, but
when the pressure is increased, a cold central core develops in addition
to the ionized interface. Fig. 1 displays the region of the cloud neutral
hydrogen column density-size plane in which a multiphase structure
exists.

\begin{figure}
\vspace{7 truecm}
\caption{ Hydrogen column density of  a HVC as a function of its size; the
numbers
refer to different values of the pressure in ergs cm$^{-3}$. {\it Solid}
triangles refer
to extragalcatic ionizing field only, {\it open}  triangles are for a composite
extragalactic + free-free form hot halo gas. The upper dashed part shows the
region in which HVC develop a multiphase structure.} \label{fig-1}
\end{figure}

Another important, though not fully understood, problem concerning
the Galactic WNM (McKee \& Ostriker 1977), is that the temperature
of this phase, as tentatively derived from observations, is "too cool''.
The problem was firstly pointed out by Heiles (1989) studying the
decomposition of the 21~cm emission line from prominent HI shells.
He argued in favour of  a component corresponding to a temperature
$\le 2000$~K.
More recently, different groups have contibuted additional evidences
along the same line, for other objects.
Dickey \etal (1990) observed the outer disk of N1058, finding a constant
velocity dispersion $\sigma\simeq 5.7$~km~s$^{-1}$ (equivalent to
$T\simeq 4000$~K) well outside the Holmberg radius. Verschuur
\& Magnani (1994), have detected a wide
component of the 21~cm line ($FWHM\simeq 12$~km~s$^{-1}$,
\ie $T\simeq 3000$~K) towards high latitude Galactic HI.

An important question to address is the nature of this "anomalous''
line broadening: is it thermal or turbulent ? Both interpretations generate
further problems.
It is well known (Field 1965) that interstellar gas in this range of
temperatures is thermally unstable. On the other hand, it is not clear what
the source of turbulence in "quiet'' regions, like outer disks and HVCs may be,
and even worse, the lack of spatial variation in $\sigma$ is hard to reconcile
with such an explanation. In order to provide some possible solutions to
the above problems we shall analyze in detail thermoreactive and acoustic
modes in a photoionized ISM and shall find a number of
new interesting physical effects which can be understood through
a comparison with
the classic Field (1965) paper. The consequences for the ISM will
have to be carefully evaluated in a future work, this
contribution is based on a paper currently submitted for publication
on ApJ (Corbelli \& Ferrara 1994, thereafter Paper I).

\section{Linear Instabilities: Acoustic and Thermoreactive Modes}
We consider an ideal homogeneous gas with metallicity $Z$,
ratio of specific heats $\gamma$, and in which hydrogen is the only reacting
species. The basic equations are
$${d\rho\over dt}+\rho{\bf \nabla\cdot}{\bf v}=0,\eqno(1)$$
$$\rho{d{\bf v}\over dt}+{\bf\nabla}p =0,\eqno(2)$$
$$N_0 {dx\over dt}- I(x,\rho,T)=0, \eqno(3)$$
$${N_0 \over \gamma -1} (1+x) k_B {dT\over dt}+N_0\Bigl({k_BT\over \gamma-1}
+\chi\Bigr){dx\over dt} + L(x,\rho,T)-{p\over \rho^2}{d\rho\over dt}=0,
\eqno(4)$$
$$p=N_0\rho (1+x) k_B T,\eqno(5)$$
where, apart from usual symbols, $x$ is the ionized fraction, $N_0$
is the Avogadro number and $\chi$ is the hydrogen ionization potential.
The equations are similar to the ones used by Field (1965); however the
reader may have noted the additional ionization equation (3). Ionization
enters also the cooling function (coupling substantially the energy and
ionization equations) and the pressure. Applying linear perturbations to
the generic equilibrium state characterized by $T=T_0, \rho=\rho_0,
v_0=0, p=p_0, x=x_0$, we obtain a {\it fourth} order dispersion relation
of the form $$\sum_{j=0}^{4}a_j n^{4-j}=0,\eqno(6) $$  where the
coefficients are given explicitly in Paper I.
Instead of the three modes found by Field (one condensation
+ 2 acoustic wave modes), we have now four modes,
almost always conjugate two by two, two of which correspond to
thermoreactive modes, and two describing acoustic waves.

In addition, the inclusion of ionization modifies the isobaric criterion
for thermal instability.
A straight consequence of Field (1965) isobaric instability criterion, is
that  thermally unstable regions are characterized
by a negative slope of the $p_0-T_0$ curve. When $L$ depends on $x$
as well as on $\rho$ and $T$, there is an additional term
proportional to the temperature derivative of the ionization fraction
$$\left({\de L\over \de T}\right)_p=\left({\de L\over \de T}\right)_{\rho,x}
-{\rho\over T}\left({\de L\over \de \rho}\right)_{T,x}+{\de x\over \de T}
\left[{\de L\over \de x}-{\de L\over \de \rho}{\rho\over 1+x}\right]_
{\rho,T}<0.$$
This criterion reduces to the Field's one when $x$ is kept constant; in all
other cases one cannot read  the stability properties directly
off the slope of the equilibrium curve.

To exemplify this important point we show in Fig. 2  different equilibrium
curves in the usual plane $p/\xi-T$. The curves have been calculated for
four different values of the metallicity in the range $Z=0,1$ and for three
values of the mean photon energy of the radiation spectrum, $E=15,
40, 100$~eV. Higher values of $E$ correspond to higher column densities,
since low energy photons have a shorter mean free path for absorption.
For every curve, regions which are linearly unstable (in the sense that at
least
one of the eigenvaues of the linearized system is positive, hence producing
instability) are denoted by dotted lines. If, for example, one applies
Field's criterion to the $Z=1, E=15$~eV curve, thermal stability would be
inferred  (since the slope
is negative), contrary to the results of our analysis.

\begin{figure}
\vspace{10 truecm}
\caption{Equilibrium curves $p_0/\xi$ for four values of the
metallicity $Z$. $p_0$ is in units of cm$^{-3}$ K and $\xi$ is the
photoionization rate in s$^{-1}$. For each value of $Z$ we show the
equilibrium curves for three energies: in each panel the bottom
curve is for $E=15$ eV, the middle curve is for $E=40$ eV and the
top curve is for $E=100$ eV.} \label{fig-2}
\end{figure}

Fig. 3 summarizes the extension of stable and unstable regions in the
$E-T$ plane (for several values of $Z$).
A distinction is also made for acoustic and thermoreactive instabilities:
dotted areas denote acoustically unstable regions, \ie only acoustic
modes are unstable for small $k$.
For low metallicity values ($Z\simlt 0.1$) such regions do not exist; they
first appear for $Z\simeq 0.5$ at low $E$. For $Z\ge 1$ two separate branches
which
are only acoustically unstable are found: one at low and one at high
temperatures and they become wider as $Z$ increases.

In the dashed regions the equilibrium
is unstable for any $k$: except close to the low and high temperature
boundaries next to the acoustically unstable regions,
the most unstable modes are the thermoreactive
ones for large $k$. Therefore we shall refer to these regions as thermally
unstable regions.  A pure hydrogen gas is thermally stable for any
temperature and photon energy, but a small amount of metals induces
the thermoreactive instability from roughly 100~K to 6000~K, even if they
grow very slowly.

\begin{figure}
\vspace{18 truecm}
\caption{Unstable regions in the plane $E-T$ for various
values of the metallicity $Z$. In the dotted regions there are
unstable modes only for $k<k_m$ (acoustic waves). In the dashed
regions equilibria have unstable modes $\forall k$
(they are both acoustically and thermoreactively unstable).
The dashed curves at the bottom and left hand side of
each panel limit the possible equilibrium values of $T_0$ for each
value of the photon energy $E$.} \label{fig-3}
\end{figure}

A major difference with respect to the standard thermal instability is
represented by the presence of unstable acoustic waves in regions
in which the gas is thermally stable. This occurence is now allowed
by the inclusion of the ionization.
Besides, especially for intermediate values of the metallicity $Z$,
thermal instability is now suppressed in some region of the parameter
spece $E,T_0$.

\section{The nature of the acoustic and thermoreactive instabilities}
In order to better understand the nature of the two instabilities (acoustic
and thermoreactive) found from the linear analysis, it is instructive to
inspect the dispersion curves in a particularly exemplificative case.
To this aim, we consider in the following the equilibrium state characterized
by a density $N_0\rho=1$~cm$^{-3}$ (wavenumber and growth rates
scale linearly with the density) and $E=100$~eV,$Z=1$. From Fig. 3
it is seen for these values, temperatures $\sim 100$~K are only
acoustically unstable for small $k$ while if $T\sim 200$~K the equilibrium
is unstable $\forall k$.
Fig. 4 shows the dispersion curves relative to this case.

\begin{figure}
\vspace{7 truecm}
\caption{Positive real part of growth rates $n_r$ as
function of the wavenumber $k$ for three equilibrium temperatures relative
to $E=100$~eV, $Z=1$, and $N_0\rho=1$~cm$^{-3}$.
For $T_0=120$ K we plot the growth rate multiplied by a factor of 10.
For this temperature the equilibrium is stable at large $k$ while for
$T_0=160$ and $T_0=200$ there is an unstable mode at all wavenumbers.
The maximum growth rate for $T_0=200$ is for $k\rightarrow\infty$.}
\label{fig-4}
\end{figure}

All three curves represent WW modes for small $k$ (\ie $k \simlt
10^{-18}$~cm$^{-1}$) and TR modes for large $k$; TR (WW) modes
in the isochoric (isobaric) regime are stable for all three cases. In the
limit $k\rightarrow 0$, WW modes are oscillatory and unstable.
For $T_0=120$~K,  $n_r$ has a maximum
at $\log k=-17.8$ followed by an abrupt break leading to stable TR modes.
The minimum growth time  for WW  is $~ 2.5\times 10^7 (N_0\rho)^{-1}$~yr
(the curve shown has been multiplied by a factor of 10).
Increasing the temperature TR modes become unstable for large $k$,
and the curve for $T_0=160$ K represents a transition case between TR
stable and unstable regimes. Except for this transition region,
when TR modes are unstable, their growth rate is usually
larger than the WW one (WW remain unstable anyway), and therefore
they represent the most dangerous modes for the system. A more typical
dispersion curve for unstable WW and TR modes is that shown for
$T_0=200$ K; as for the standard thermal instability $n_r$ grows
monotonically with $k$ and the curve flattens out as
$k\rightarrow\infty$. The effect of thermal conduction in this case
would be to stabilize very large wavenumber perturbations ($k>k_c=
10^{-15}$~cm${-1}$, for the parameters considered).

The physical nature of the acoustic instability can be understood as follows.
When a fluid element is compressed and rarefied by a travelling wave,
there will be a net energy flow from
the gas to the wave if $\gamma$ increases from the compression to the
expansion phase. If the ionization time is longer than the cooling time,
compression will
proceed at constant $x$ and almost isothermally if the gas is stable,
since this requires a steep dependence of the cooling function on $T$.
Suppose that $L$ depends
on $x$ in such a way that a lower $x$ requires a higher $T$ to provide
the same cooling; on the ionization time scale,
$x$ will start to decrease because of the higher density,
temperature (and hence pressure) will increase destabilizing the wave.
In other words, the ionization provides an effective increase of $\gamma$.
As a consequence, the existence of the critical wavenumber
for WW stabilization is fixed by the condition that the dynamical
time is roughly equal to the ionization time.
At that value of $k$ the oscillation rate ($\propto k$ in the
entire range) becomes
much shorter than the ionization time and stiffening
of the fluid caused by ionization is no longer possible.

For $T_0=120$ K Figure 4 shows that the system is acoustically unstable
but thermoreactively stable;
note, however, that without having considered the effects of ionization,
it would undergo already to thermal instability at this temperature.
For the equilibrium conditions we have considered
{\it ionization is in fact a stabilizing agent} with respect to thermal
instability when it introduces and oscillatory part in the growth rate.
This oscillatory character often shown by thermoreactive instabilities
can be illustrated with simple arguments.
Suppose that the cooling is dominated by electron-metal impacts
and to decrease slightly the temperature. Since TR modes are essentially
isobaric, the density will then increase, and, if the initial state is
thermally unstable, $T$ will increase even further.
However, on the ionization time scale (if not too large)
the system will react decreasing $x$.
If $L_x>0$ (at constant pressure),  the cooling will decrease,
and as a consequence
the temperature will raise back again, hence counteracting the instability
and producing the unstable oscillatory motion (overstability).
The previous condition  $L_x>0$ is always verified
when cooling is dominated by electron-metal impact and $x\ll1$,
since in that case the heat-loss function depends linearly on $x$.
An increase of the metallicity $Z$ in this unstable region rises
the hydrogen fractional ionization and lowers the reactive
times allowing the TR modes to become oscillatory.

The overstability of the TR modes produces interesting effects in
the nonlinear stages, which are the subject of the next Section.

\section{Nonlinear evolution of TR modes}

The aim of this Section is to investigate the nonlinear evolution of
perturbations imposed to equilibrium states which are found to
be thermoreactively unstable from the linear analysis.
Since the results depend crucially on the phase structure of
the gas, we classify the equilibria in two families:
unstable multi-phase
(\ie two or more isobaric equilibria of which
at least one is stable) or unstable single-phase (\ie one or more isobaric
unstable equilibria). When the photon energy is high ($E\gg 13.6$~eV)
unstable equilibria are multi-phase; in the opposite case they are always
single-phase (see Fig. 2). We will discuss the results using the equilibrium
curve corresponding to $E=40$~eV, $Z=1$, which shows both equilibrium
states depending on the value of $p_0/\xi$. The curve is reported in the top
panel of Fig. 5 and has been expanded for convenience; solid (dashed)
lines denote thermoreactively unstable (stable) equilibria.
Each of the bottom panels shows
the nonlinear evolution of perturbations (with amplitude $\sim 5$\%) imposed
to the four equilibrium points P1-P4.

\begin{figure}
\vspace{18 truecm}
\caption{{\it Top:} Expanded equilibrium curve $p_0/\xi-T_0$
for $E=40$ eV and $Z=1$; dashed-dotted lines refer to
thermoreactively unstable equilibria; P1-P4 mark points
for which nonlinear evolution is shown below.
{\it Bottom:} Nonlinear evolution of the
temperature and ionization fraction perturbations in the isobaric
regime.
Continuous curves refer to $\delta T/T_0$, dotted curve to
$\delta x/x_0$ as function of time (normalized to $\tau_H$).
The density is set to $N_0\rho=1$ cm$^{-3}$.} \label{fig-5}
\end{figure}


The first point $P1$ is a thermoreactively stable equilibrium with
$T_0=120, x_0=0.012$~K. In the linear phase the perturbation shows
oscillations with $n_i=5\times 10^{-6}$~yr$^{-1}$. These oscillations are
clearly seen also during the nonlinear evolution before the perturbation
is damped on a timescale  $\sim 400 \tau_h=10^7$~yr, where
$\tau_h\equiv (5/2) kT/\xi E$ is the local heating time.

$\bullet$ {\it Multi-phase equilibria}. The unstable equilibrium P4
corresponds to $T_0=2700, x_0=0.18$ and is obtained
when $\xi\simeq 3\times 10^{-14}$~s$^{-1}$ for $N_0\rho=1$~cm$^{-3}$.
The same value of $p_0/\xi$ corresponds to the stable equilibrium point
$T_0=8100, x_0=0.44$ with a value of the volume density about 3 times
smaller. In the linear phase $P4$ is thermoreactively unstable with
$n_i=0$. The system reaches the other stable equilibrium after about
600$\tau_h=7.5\times 10^6$~yr through (nonlinear)
oscillations. This time is much longer than the typical time scale
that one would have calculated without the ionization, and, indeed, the
system spends a long time in a time-dependent, nonequilibrium state.
This retarding effect, due to ionization and already
discussed in the previous Section, occurs very often.

$\bullet$ {\it Single-phase equilibria}. Equilibria $P2$ and $P3$ in Fig.
5 correspond to single-phase equilibria.
For $P2$, $T_0=150$~K and the perturbation is  overstable in the linear
phase with an oscillation rate larger than
the growth rate. For this reason, a large number of oscillations of growing
amplitude is seen in $T,x,\rho$ before saturation takes place.
After $t\sim 10^3\tau_h$ the system relaxes to a state of {\it periodic}
oscillations of constant amplitude. Note, that the ionization lags the
temperature, as expected.
Far from the unstable/stable transition region  $n_i$ is zero
or rather small compared to the to the growth rate.
Therefore for $P3$ the perturbations grow to the saturated values in less
than one oscillation period. After that moment, the evolution is
qualitatively similar to that of $P2$ but the
temperature spans the very large range $1000\simlt T \simlt 10^4$~K,
since the oscillation is not centered around $T_0$.

To conclude, isobaric TR instabilities produce much slower phase
transitions than the usual thermal instabilities. When no other
stable equilibrium is available the medium spends long time
intervals in  nonequilibrium states characterized by nonlinear
periodic oscillations of temperature, density and ionization fraction.

\section{Summary}

We have studied acoustic and thermoreactive instabilities
in a diffuse
ISM heated and ionized by and external radiation field, of mean photon
energy $E$ and photoionization rate $\xi$, cooled
by collisional excitation of metals and hydrogen lines. The addition of
the hydrogen recombination reaction with respect to Field's classical
treatment of thermal instability has several remarkable consequences
especially when
the ratio between the heat input and the ionization rate of the radiation
field is rather small. In addition to the linear stability analysis we
have investigated the nonlinear development of thermoreactive modes.

In the absence of metals the gas is stable.
A very small amount of metals induces unstable
acoustic and nonoscillatory thermoreactive modes and, as in
Field's analysis, the fastest growing instability is the isobaric
thermal mode.
If the metal abundance is about half solar or higher the linear analysis
shows that unstable modes often become oscillatory and ionization is
a stabilizing agent, as explained in the  previous Sections.
Futhermore, a new type of situation arises in which the
gas is thermoreactively stable but unstable acoustic waves develop on
large scales. While for $Z\simeq 0.5$ this occurs only for
optically thin media, with $E-\chi \ll 100$ eV, for larger $Z$ two
regions are found which are only acoustically unstable, one in the cold
phase at $T\sim 100$~K and one in the warm phase at $T\sim 8000$~K.

Another new and important feature introduced by the hydrogen
recombination reaction
is that the stability of isobaric thermoreactive modes is no longer
connected with the shape of the $p_0/\xi$ curve. This implies that there
are unstable equilibria which cannot make an isobaric phase transition to
any other stable state. The nonlinear analysis has shown
that in this case the fate of the gas is to evolve towards a nonequilibrium
state characterized by periodic, nonlinear oscillations of density,
temperature and hydrogen ionization fraction.
This effect is especially
important for small heat input/ionization rate ratios,
and for $Z>0.1$, but in general
it would be interesting to extend this type of  analysis to additional
ionization and heating mechanisms (decaying neutrinos, dust grains).

A number of possible applications to astrophysical objects are presented
in Paper I.

\acknowledgments
This work has benefited by a large number of inspiring and stimulating
discussions with Prof. George B. Field. Daily contact with him has
been a source of scientific and cultural enrichment, but above
all, a unique human experience.

\end{document}